\documentclass[aps,prl,twocolumn,showpacs,superscriptaddress,reprint]{revtex4-1}
\usepackage{graphicx}
\usepackage{natbib}
\usepackage{amsmath}
\usepackage{amssymb}
\usepackage{bm}
\usepackage{color}

\begin{document}

\title{Coexistence of coupled magnetic phases in epitaxial TbMnO$_3$ films revealed by ultrafast optical spectroscopy}

\author{J. Qi} 
\email{jingbo@lanl.gov}
\affiliation{Center for Integrated Nanotechnologies, Los Alamos National Laboratory, Los Alamos, NM 87545}
\author{L. Yan} 
\affiliation{Center for Integrated Nanotechnologies, Los Alamos National Laboratory, Los Alamos, NM 87545}
\author{H. D. Zhou}
\affiliation{Department of Physics and Astronomy, University of Tennessee, Knoxville, TN 37996} 
\affiliation{National High Magnetic Field Laboratory, Tallahassee, FL 32306}
\author{J.-X. Zhu} 
\affiliation{Theoretical Division, Los Alamos National Laboratory, Los Alamos, NM 87545}
\author{S. A. Trugman} 
\affiliation{Center for Integrated Nanotechnologies, Los Alamos National Laboratory, Los Alamos, NM 87545}
\affiliation{Theoretical Division, Los Alamos National Laboratory, Los Alamos, NM 87545}
\author{A. J. Taylor} 
\affiliation{Center for Integrated Nanotechnologies, Los Alamos National Laboratory, Los Alamos, NM 87545}
\author{Q. X. Jia} 
\affiliation{Center for Integrated Nanotechnologies, Los Alamos National Laboratory, Los Alamos, NM 87545}
\author{R. P. Prasankumar}
\email{rpprasan@lanl.gov} 
\affiliation{Center for Integrated Nanotechnologies, Los Alamos National Laboratory, Los Alamos, NM 87545}


\begin{abstract}
Ultrafast optical pump-probe spectroscopy is used to reveal the coexistence of coupled antiferromagnetic/ferroelectric and ferromagnetic orders in multiferroic TbMnO$_3$ films through their time domain signatures. Our observations are explained by a theoretical model describing the coupling between reservoirs with different magnetic properties. These results can guide researchers in creating unique kinds of multiferroic materials that combine coupled ferromagnetic, antiferromagnetic and ferroelectric properties in one compound. 
\end{abstract}

\maketitle
Multiferroic materials have attracted much interest in the past decade, due not only to their potential device applications, but also their manifestations of coupling and interactions between different order parameters (particularly electric polarization and magnetic order). Recently, much attention has been focused on perovskite manganites, $R$MnO$_3$ ($R$=rare earth ions), due to the discovery of a large magnetoelectric effect in these materials~\cite{Kimura_2007_AnRMR, Cheong_2007_NatMa}. The first member of this family to be discovered was TbMnO$_3$ (TMO), which is now well established as a typical magnetoelectric multiferroic.

Extensive studies have already been done on single crystal TMO (SC-TMO) \cite{Kimura_2007_AnRMR, Cheong_2007_NatMa}. In brief,  SC-TMO, with a distorted orthorhombic perovskite structure, has an antiferromagnetic (AFM) phase transition at $T_N\simeq40$ K with sinusoidally ordered Mn moments. Below $T_{FE}\simeq28$ K, ferroelectric (FE) order develops owing to the appearance of cycloidal spiral AFM spin structure. In contrast, there are relatively few reports describing the properties of TMO thin films (typically grown on SrTiO$_3$ (STO) substrates) \cite{Cui_2005_SSCom, Rubi_2009_PRB, Kirby_2009_JAP, Venkatesan_2009_PRB, Daumont_2009_JPCM, Marti_2010_APL}. In general, thin films can enable enhanced functionality in materials, as their physical parameters can be changed by modifying their structure via strain imposed by the substrate. Strain, in particular, can possibly provide a different route to direct FE-FM coupling, which is very rare \cite{Cheong_2007_NatMa, Ramesh_2007_NatMa}. This could benefit electronic device applications by providing low power consumption, high speed operation, and greater electric/magnetic field controllability \cite{Ramesh_2007_NatMa}. 

In TMO films, previous investigations of magnetic properties revealed an unexpected ferromagnetic (FM) order \cite{Rubi_2009_PRB, Kirby_2009_JAP, Marti_2010_APL}, in contrast to SC-TMO. However, none of these reports have demonstrated the existence of FE (or equivalently, spiral AFM) order in these thin films. Furthermore, one can ask: if FE order does exist, is it coupled to FM order? 

Here, we use ultrafast optical pump-probe spectroscopy to reveal coexisting coupled magnetic and ferroelectric orders in epitaxially grown TMO thin films, which were not observed in previous work \cite{Cui_2005_SSCom, Rubi_2009_PRB, Kirby_2009_JAP, Venkatesan_2009_PRB, Daumont_2009_JPCM, Marti_2010_APL}. Our temperature (\textit{T})-dependent transient differential reflectivity ($\Delta R/R$) measurements show clear signatures of FE and AFM phases developing as the film thickness changes. We propose a model to explain the interactions between AFM/FE and FM orders. These results reveal that the coupling between different magnetic and FE orders observed in our TMO thin films may offer greater control of functionality as compared to bulk multiferroics.    

Our transient differential reflectivity experiments used a Ti:Sapphire laser oscillator with a repetition rate of 80 MHz producing $\sim$140 femtosecond (fs) pulses at the sample position, with a center wavelength of $\sim$830 nm (1.5 eV). The pump and probe pulses are cross-polarized and normally incident on the sample surface. The pump has a typical fluence of 1 $\mu J$/cm$^2$, which is 10 times larger than that of the probe. This excites $\sim10^{-4}$ quasiparticles/unit cell on average, ensuring that the photoexcitation does not induce large thermal or non-thermal effects. 

The TMO films investigated here were grown on STO(001) substrates using pulsed laser deposition (PLD), with thicknesses of $d$=34, 50, and 150 nm. We also examined a TMO single crystal (SC-TMO) grown through the floating zone technique and a 50 nm SmMnO$_3$ (SMO) film grown by PLD on a STO (001) substrate. In the bulk, SMO displays A-type AFM order below $\sim60$ K, but unlike TMO, does not exhibit sinusoidal/cycloidal spin order or FE order \cite{Kimura_2003_PRB}. Therefore, comparing measurements on our TMO films with those in a TMO single crystal and a SMO film will help us unravel the influence of these order parameters on quasiparticle dynamics in TMO. 

\begin{figure}[h]
\includegraphics[width=7cm]{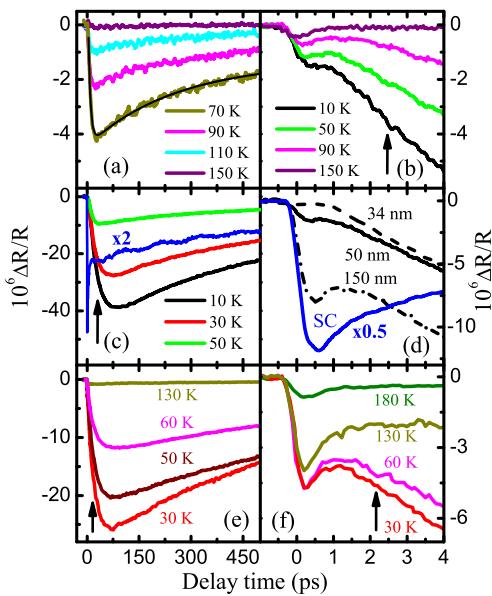}
\caption{\label{fig:deltaR} $\Delta R/R$ signals measured on a 50 nm TMO film on long ((a) and (c)) and short timescales (b). Part (d) shows ultrafast dynamics at 20 K on short timescales in three TMO films with $d\sim$34 nm (dashed), 50 nm (solid), and 150 nm (dashed-dotted), respectively. The black solid curve in (a) demonstrates the fitting. For comparison, data at 20 K for SC-TMO (blue) is also shown in (c) and (d). $\Delta R/R$ signals measured on a 50 nm SMO film on long and short timescales are shown in (e) and (f), respectively.}
\end{figure}

Figures \ref{fig:deltaR}(a), (b) and (c) show typical $\Delta R/R$ data obtained on a $d=$50 nm TMO film. Very similar results are obtained on the 34 and 150 nm films, except that the initial signal magnitude increases with thickness (Fig. \ref{fig:deltaR}(d)).  The dynamics measured in the SMO film (Figs. \ref{fig:deltaR}(e) and (f)) are also quite similar to those measured in the TMO films. We observe three distinct relaxation processes in these films, occurring on different timescales. Within a time $t\sim$1.5 picoseconds (ps), there is an ultrafast increase in the magnitude of the reflectivity, followed by a fast signal reversal. The former is due to the photoexcitation of carriers via intersite Mn \textit{d-d} transitions, which dominate the optical response at 1.5 eV in perovskite manganites \cite{Kovaleva_2004_PRL, Bielecki_2010_PRB}. We note that the strong temperature dependence observed in our data further supports this, and indicates that \textit{p-d} transitions do not play a significant role \cite{Bielecki_2010_PRB}. The fast signal reversal, occurring at $t\sim1$ ps, can be mainly attributed to electron-phonon (e-ph) relaxation, as in other manganites \cite{Wall_2009_PRL}. The timescale of these initial ultrafast changes is approximately independent of the temperature and film thickness. 

After the initial ultrafast dynamics are complete, an additional rising process (in which the magnitude of $\Delta R/R$ increases, indicated by the arrows in Fig. \ref{fig:deltaR}), clearly appears in our films around $\sim$1.5 ps below a temperature $T_C$ ($\sim$140 K) followed by a much slower relaxation process, with a decay time of hundreds of picoseconds, that returns the photoexcited system back to equilibrium. The latter process is likely due to heat diffusion, as will be discussed in more detail below. The rising process, occurring on an intermediate time scale of tens of ps at low temperatures (\textit{T}), is unlikely to be due to long-range bulk-like AFM or FE order (as one would expect in SC-TMO) for two reasons. First, the transition temperature for the emergence of this additional rising process, $T_C$, is well above $T_N$ and $T_{FE}$ of SC-TMO (Fig. \ref{fig:deltaR}(b)). Secondly, our data reveals that this process is absent in SC-TMO (Figs.~\ref{fig:deltaR}(c) and (d)). Instead, we only found two relaxation processes in SC-TMO, which have timescales of $\sim$1 ps and hundreds of picoseconds, respectively. The short-lived ps dynamics in SC-TMO, representing rapid thermalization of the excited charge, lattice and spin subsystems, are consistent with previous measurements on other long-range AFM ordered compounds \cite{Wall_2009_PRL, Satoh_2007_PRB, Tamaru_2008_PRB, Kirilyuk_2010_RMP}, and the slower dynamics are due to thermal diffusion \cite{Satoh_2007_PRB}. The rising process observed in our TMO (and SMO) films is thus unlikely to originate solely from the properties of their bulk counterparts. Instead, we note that a relatively slow spin-lattice relaxation process associated with long-range FM order was previously observed in some perovskite oxides \cite{Lobad_2000_APL, Ogasawara_2005_PRL}. Therefore, the rising process observed here might be linked to the presence of FM order in our films.  

\begin{figure}[h]
\includegraphics[width=8cm]{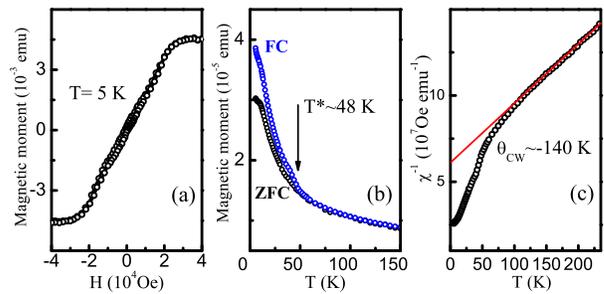}
\caption{\label{fig:SQUID} (a) The magnetic moment versus magnetic field H at $T$ = 5 K is shown for a 50 nm TMO film. ZFC-FC curves are shown in (b). The inverse susceptibility ($\chi^{-1}$) of the film is plotted as a function of temperature, $T$, in (c). The data was taken with a 1 kOe field applied parallel to the sample plane.}
\end{figure}

To investigate this possibility, we measured the field (\textit{H})-dependent magnetic moment in our films using a SQUID magnetometer. Figure \ref{fig:SQUID}(a) shows hysteretic behaviour in the 50 nm TMO film, confirming the presence of ferromagnetism. This is supported by temperature-dependent measurements (Fig. \ref{fig:SQUID}(b)), where a clear splitting of field-cooled (FC) and zero-field-cooled (ZFC) curves below $T^*\simeq$48 K reveals the appearance of a net magnetization. To uncover more information on the magnetic order(s), we plotted the inverse susceptibility $\chi^{-1}$ as a function of temperature (Fig. \ref{fig:SQUID}(c)). We did not observe clear anomalies at $\chi^{-1}(T_{FE})$ in our TMO films. By fitting the data in the high-temperature regime with the Curie-Weiss form $\chi^{-1}\propto(T-\theta_{CW})$, we found the Curie-Weiss temperature, $\theta_{CW}\simeq-140$ K. The negative $\theta_{CW}$ indicates that the exchange interactions in our films are primarily antiferromagnetic \cite{Ramirez_1994_ARMS}. Similar trends were observed in our other films, consistent with previous results \cite{Rubi_2009_PRB, Kirby_2009_JAP, Marti_2010_APL}. However, we note that the exact mechanism causing the observed ferromagnetism is still unclear. For example, ferromagnetism could arise from a strain-induced change in the balance between the different magnetic exchange interactions \cite{Rubi_2009_PRB, Marti_2010_APL}, electronic phase separation (due to nucleation of a ferromagnetic phase that is induced by epitaxial strain \cite{Moskvin_2009_PRB}), or the presence of twin domain walls due to film strain relaxation \cite{Daumont_2010_arXiv}.  

\begin{figure}[h]
\includegraphics[width=7cm]{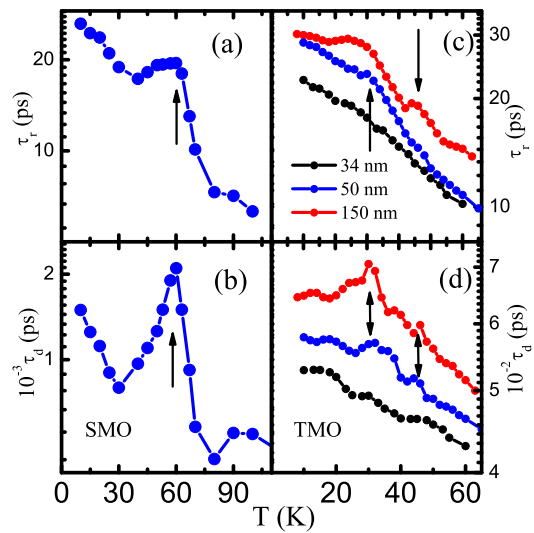}
\caption{\label{fig:constants} Decay time constants ($\tau_r$ and $\tau_{d}$) at different temperatures for the $d$=50 nm SMO film ((a) and (b)), and for $d$=34 nm, 50 nm, and 150 nm TMO films ((c) and (d)). For clarity, $\tau_r$ for the 50 nm and 150 nm TMO films in (c) is shifted up by 3 ps and 6 ps, respectively. $\tau_d$ for the 50 nm and 150 nm TMO films in (d) is shifted up by 50 ps and 150 ps, respectively. The arrows indicate the phase transition temperatures.}
\end{figure}

Our magnetization measurements thus indicate that the rising process is linked to the presence of FM order in our films; however, this process emerges at $T_C$, well above $T^*$, suggesting that additional physics could be involved. For more insight, we fitted our data for $t\gtrsim 1$ ps with the equation: $A_re^{-t/\tau_r}+A_de^{-t/\tau_d}$ ($A_r>0$ and $A_d<0$) (Fig. \ref{fig:deltaR}(a)). This revealed some surprising anomalies in the temperature variation of the intermediate ($\tau_{r}$) and slow ($\tau_{d}$) time constants for the SMO and TMO films studied here (Figure \ref{fig:constants}) that were not observed in the SQUID measurements. 

In particular, very strong peaks in both $\tau_{r}(T)$ and $\tau_{d}(T)$ are found in the SMO film at $\sim$60 K. As discussed above, SC-SMO has an A-type magnetic structure below the N$\acute{e}e$l temperature ($T_N\simeq60$ K). Therefore, the anomalies in the measured time constants at $\sim$60 K should be associated with the appearance of long-range bulk-like AFM order. Similarly, in our TMO films, both $\tau_{r}$ and $\tau_{d}$ show clear kinks at $\sim$30 K and $\sim$46 K for $d$=50 nm and 150 nm (most obvious in the latter), as indicated by the arrows. If either of these anomalies is primarily due to FM order, we also expect to observe it in the 34 nm TMO film based on the SQUID results and strain-related ferromagnetism. However, no obvious anomalies were found for the 34 nm film. Thus, as in SMO, since 30 K and 46 K are very close to $T_{FE}$ and $T_{N}$ in SC-TMO, the anomalies in $\tau_{r}$ and $\tau_{d}$ can be associated with the bulk-like FE and AFM phase transitions, respectively. In addition, the fact that 46 K is so close to $T^*$ suggests that TMO films may undergo AFM and FM transitions at the same temperature.   

It is relatively straightforward to explain the anomalies in the slower time constant, $\tau_d$. This process originates from thermal diffusion, including both lateral diffusion within the films as well as diffusion into the substrate (the former becomes more important with increasing film thickness). This can be shown by considering the diffusion equation ${\partial T}/{\partial t}=\alpha\nabla^2 T$, in which the heat transfer rate, reflected in $1/\tau_{d}$, decreases with the thermal diffusivity $\alpha$ ($=\kappa/C\rho$), where $\kappa$, $C$ and $\rho$ are the heat conductivity, the specific heat and the density, respectively. Thus, when $\kappa^{-1}$ and $C$ peak at the magnetic phase transition temperatures for TMO and SMO \cite{Kimura_2003_PRB, Berggold_2007_PRB}, thermal diffusion will correspondingly slow down, explaining the anomalous peaks in $\tau_{d}$.    
 
The origin of the intermediate time constant $\tau_r$ requires more detailed consideration. In all TMO films studied here, we found $T_C\simeq|\theta_{CW}|$, suggesting that not only FM ordering, but also AFM exchange interactions, may influence this rising process. This could occur through strong magnetic frustration due to competing magnetic interactions, which is often observed in multiferroics \cite{Cheong_2007_NatMa}. One signature of this is when $|\theta_{CW}|$ is much greater than $T^*$ and $T_N$ \cite{Cheong_2007_NatMa, Ramirez_1994_ARMS}, which is the case for our films (Fig. \ref{fig:SQUID}). This is further supported by our ultrafast optical data, where clear anomalies associated with other magnetic orders are found in $\tau_r$ at $T_N$ and $T_{FE}$, indicating that they coexist with the FM order revealed by our SQUID measurements. The magnetic interactions inducing the long-range bulk-like AFM/FE order may thus compete with the magnetic interactions contributing to ferromagnetism, influencing quasiparticle dynamics on the timescale of $\tau_r$.

Motivated by these findings, we propose a model with two spatially separated magnetic reservoirs, one with bulk-like sinusoidal/spiral AFM order and the other with FM order, to describe the intermediate rising dynamics observed in $R$MnO$_3$ films. They are characterized by two effective temperatures, $T^{A}$ and $T^{F}$, respectively. The coupling between the two reservoirs occurs via the exchange spin-spin interaction $J_{A-F}{\bf S}^A\cdot {\bf S}^F$ between these two systems \cite{Frobrich_2006_PhysRep}. This coupling in the photoexcited system is manifested through energy transfer via magnon-magnon coupling/conversion between the two reservoirs, where the magnon excitations in both reservoirs are introduced through intersite \textit{d-d} transitions \cite{Wall_2009_PRL}. Similar to the classic two-temperature model describing electron-phonon coupling \cite{Averitt_2002_JPCM}, the dynamical behavior of the photoexcited coupled system can be described by the time evolution of its effective temperatures, given by
\begin{align}
C^{A} \dfrac{d}{dt}(T^{A})=&g_{A-F}(T^{F}-T^{A}),
\nonumber\\
C^{F}\dfrac{d}{dt}(T^{F})=&g_{A-F}(T^{A} -T^{F})
\label{eq:TTM},
\end{align}
where $g_{A-F}$ is the energy transfer rate, and $C^{j}$ ($j=A, F$) are the magnetic heat capacities for the two reservoirs with AFM order and FM order, respectively. After solving these differential equations, we found that the non-equilibrium coupled system can be characterized by a time constant $\tau$, 
\begin{align}
\tau^{-1}=g_{A-F}\left(\frac{1}{C^{A}}+\frac{1}{C^{F}}\right)
\label{eq:tau_TTM}.
\end{align} 
Now, based on this theoretical $\tau$ that describes the intermediate rising process, we can understand how the magnetic phase transitions could be revealed in the temperature dependence of $\tau_r$ (Fig. \ref{fig:constants}). Specifically, $\tau$ will reach a maximum in the vicinity of a critical temperature because: (1) the heat capacities $C^A$ and $C^F$ peak at critical temperatures \cite{Kimura_2003_PRB}, and (2) $g_{A-F}$ might become smaller at the magnetic phase transition temperatures due to the \textit{critical slowing down} phenomenon \cite{Domb_1972_Academic}. Furthermore, if we assume $C^{A}$ and $C^{F}$ have similar values, we can estimate the coupling constant $g_{A-F}\simeq2\times10^{10}$ WK$^{-1}$mole$^{-1}$ for TMO films at $30$ K, calculated by taking the spin specific heat $C^{A}\simeq1$ JK$^{-1}$mole$^{-1}$ and  $\tau_r\simeq25$ ps \cite{Kimura_2003_PRB,Choithrani_2011_JMMM}. Our model thus shows that the anomalous peaks in $\tau_r(T)$ are due to the presence of coexisting, coupled magnetic orders in these films. 

\begin{figure}[h]
\includegraphics[width=8cm]{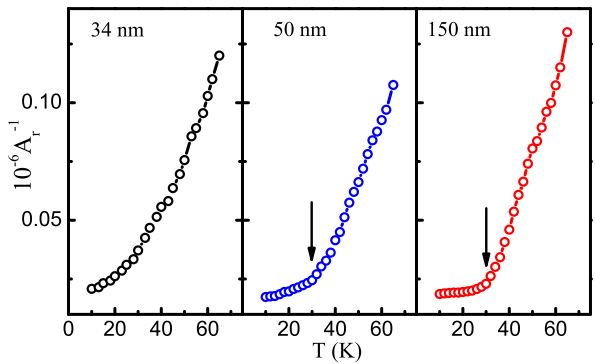}
\caption{\label{fig:amplitude} Temperature dependence of $A_r^{-1}$ for three TMO films.}
\end{figure}

It follows that the appearance of the intermediate rising process ($A_r>0$) below $|\theta_{CW}|$ represents the competition between magnetic interactions. Thus, the amplitude $A_r$ reflects the coupling between neighboring spins, expressed by the spin-spin correlation function $<S_iS_j>$, which includes intra-reservoir correlations between neighboring spins in each reservoir as well as the inter-reservoir correlation function $<S^A_iS^F_j>$; this last term also contributes to the coupling constant $g_{A-F}$ in Eq. (\ref{eq:TTM}). Figure \ref{fig:amplitude} demonstrates that $A_r^{-1}$ decreases with decreasing $T$. This may indicate that $<S_iS_j>$ increases as the temperature decreases, as in previous work \cite{Kovaleva_2004_PRL}. Most interestingly, we found a very clear transition at $T\sim30$ K, comparable to the bulk $T_{FE}$, for both 50 nm and 150 nm TMO films. Below $\sim30$ K, $A_r^{-1}$ is nearly constant with temperature, unlike its behavior for $T>30$ K. Since $A_r(T)$ is influenced by the spin-spin correlation function $<S_iS_j>$, this result is consistent with previous work showing that the propagation spin wave vector {\bf\it k}$_s$ is locked at a constant value below $T_{FE}$ ($\sim$30 K) \cite{Kimura_2007_AnRMR} when TMO undergoes a FE phase transition and the spin structure changes from sinusoidal to spiral; this further supports the existence of FE order in our TMO films.    

In conclusion, we have used ultrafast optical pump-probe spectroscopy to reveal coexisting, coupled magnetic and ferroelectric orders in epitaxial TMO films through their time domain signatures. This could lead to an enhanced magnetoelectric effect in these films as compared to their bulk counterparts, which would allow the FE polarization and/or magnetization in TMO films to be more easily controlled by external fields than in bulk TMO.  Our results therefore indicate that thin perovskite manganite films could be an unique kind of multiferroic combining coupled FM, AFM and FE properties in a single material.

This research was performed at the Center for Integrated Nanotechnologies, a U. S. Department of Energy, Office of Basic Energy Sciences user facility, and supported by the Laboratory Directed Research and Development Program. H. D. Zhou is supported by NSF-DMR-0654118 and the State of Florida. We acknowledge useful discussions with A. V. Balatsky.

\end{document}